\documentstyle[11pt,aas2pp4,epsfig]{article}
\begin{document}

\title{On the Amplitude of Burst Oscillations in \\ 4U 1636-54: Evidence for
Nuclear Powered Pulsars}
\author{Tod E. Strohmayer\altaffilmark{1},  William Zhang\altaffilmark{3}, Jean
H. Swank\altaffilmark{3}, Nicholas E. White\altaffilmark{3}}
\affil{Laboratory for High Energy Astrophysics \\ Goddard Space Flight Center \\
Greenbelt, MD 20771}
\author{Iosif Lapidus\altaffilmark{2}}
\altaffiltext{1}{ {\it also} USRA/LHEA, Mail Code 662, NASA/GSFC Greenbelt, MD
20771}
\affil{Univeristy of Sussex, England}
\altaffiltext{2}{University of Sussex, England}
\altaffiltext{3}{LHEA, Mail Code 662, NASA/GSFC Greenbelt, MD 20771}
\authoraddr{Laboratory for High Energy Astrophysics, Mail Code 662, NASA/GSFC
Greenbelt, MD 20771}

\begin{abstract}

We present a study of 581 Hz oscillations observed
during a thermonuclear X-ray burst from the low mass X-ray binary (LMXB)
4U 1636-54 with the Rossi X-ray Timing Explorer (RXTE). This is the first
X-ray burst to exhibit both millisecond oscillations during the rising phase
as well as photospheric radius expansion. We measure an oscillation
amplitude within 0.1 s of the onset of this burst of $75 \pm 17 \%$, that is,
almost the entire thermal burst flux is modulated near onset. The spectral
evolution during the rising phase of this burst suggests that the X-ray emitting
area on the neutron star was increasing, similar to the behavior of bursts from
4U 1728-34 with 363 Hz oscillations reported recently. We argue that the
combination of large pulsed amplitudes near burst onset and the spectral
evidence for localized emission during the rise strongly
supports rotational modulation as the mechanism for the oscillations. 
We discuss how theoretical interpretation of spin modulation amplitudes, 
pulse profiles and pulse phase spectroscopy can provide constraints on the
masses and radii of neutron stars. We also discuss the implications of these
findings for the beat frequency models of kHz X-ray variability in LMXB.

\end{abstract}


\section{Introduction}

Large amplitude millisecond oscillations have now been observed during
thermonuclear X-ray bursts from six low mass X-ray binary (LMXB) systems with
the Rossi X-ray Timing Explorer (RXTE) (see \markcite{SZS}Strohmayer, Zhang \&
Swank 1997; \markcite{SMB}Smith, Morgan \& Bradt 1997, \markcite{Z96}Zhang {\it
et al.} 1996; \markcite{Swank97}Swank {\it et al.} 1997;
and \markcite{Stroh97}Strohmayer {\it et al.} 1997). The thermonuclear
instability which triggers an X-ray burst burns in a few seconds the nuclear
fuel which has been accumulated on the neutron star surface over several hours.
This $>$ 10$^3$ difference between the accumulation and burning timescales means
that it is extremely unlikely that the conditions required to trigger the
instability will be achieved simultaneously over the entire stellar surface.
This realization, first emphasized by \markcite{Joss78}Joss (1978), led to the
study of lateral propagation of the burning instability over the neutron star
surface (see \markcite{FW82}Fryxell \& Woosley 1982, \markcite{NIF84}Nozakura,
Ikeuchi \& Fujimoto 1984, and \markcite{B95}Bildsten 1995). The subsecond
risetimes of thermonuclear X-ray bursts suggests that
convection plays an important role in the physics of the burning front
propagation, especially in the low accretion rate regime which leads to large
ignition columns (see \markcite{B98}Bildsten (1998) for a review of
thermonuclear burning on neutron stars). \markcite{B95}Bildsten (1995) has shown
that pure helium burning on neutron star surfaces is in general inhomogeneous,
displaying a range of behavior which depends on the local accretion rate. 
Low accretion rates lead to convectively
combustible accretion columns and standard type I bursts, while high accretion
rates lead to slower, nonconvective propagation which may be manifested in hour
long flares. These studies emphasize that the physics of thermonuclear burning
is necessarily a multi-dimensional problem and that {\it localized} burning is
to be expected, especially at the onset of bursts. 

There is now good evidence that the oscillations seen during the rising phase of
bursts from  4U 1728-34 are produced by spin modulation of such a localized
thermonuclear hotspot on the surface of the neutron star, and that the observed
oscillation frequency is a direct measure of the neutron star spin frequency
(see \markcite{SZS}Strohmayer, Zhang \& Swank 1997). These observations provide
the most compelling evidence to date that neutron stars in LMXB are rotating
with near millisecond periods. 

In this Letter we present new burst data from the LMXB 4U 1636-54 which
provides further evidence in support of the spin modulation hypothesis
for the millisecond burst oscillations. We present data from a thermonuclear
burst from 4U 1636-54 which reveals a
strong transient oscillation during the burst rise at 1.723 ms with an initial
amplitude of $75 \pm 17 \%$. We also discuss the implications of 
these findings for the spin modulation interpretation and how they can be used
to place constraints on the mass and radius of neutron stars. Finally, we
discuss some implications of our observations for the current theories of
kilohertz quasiperiodic oscillations (QPO) in LMXB.
 
\section{Observations and data description}

\markcite{Z96}Zhang {\it et al.} (1996) reported the discovery of 581 Hz
oscillations during
several thermonuclear X-ray bursts from 4U 1636-54. Here we focus on 
a single burst detected from this source with RXTE at 22:39:22 UTC on December
28, 1996. This burst is unique in that it reveals strong oscillations at 
581 Hz {\it prior} to the onset of photospheric radius expansion.
Oscillations are not detected during the photospheric 
expansion phase, but reappear with much weaker amplitude after photospheric
touchdown in a manner very similar to that reported for bursts from 4U1728-34
and KS 1731-26 (\markcite{Stroh97}Strohmayer {\it et al.} 1997; and
\markcite{SMB}Smith, Morgan, \& Bradt 1997). 

For this
burst we obtained event mode data with 125 $\mu$s (1/8192 s) time resolution and
64 spectral channels across the
2-100 keV PCA bandpass. This data mode employs a pair of data buffers which are
alternately read out after a selectable accumulation interval. Each buffer can
store $\approx$ 16,000 events. In the case that the source countrate is high
enough to fill the buffer before the end of an accumulation interval gaps in the
data can occur. For the event data shown here we did have short gaps when the
burst flux was high, but telemetry constraints preclude selecting a shorter
accumulation interval. For these bursts we also had 125 $\mu$s, 2-90 keV binnned
timing data which does not suffer from gaps.

\section{Oscillations during the rise of a radius expansion burst}

At the time of the burst in question the persistent countrate from 4U1636-54 
in the total PCA bandpass was about 2100 cts/sec. 4U1636-54 is classified as
an atoll source and it was in the so called banana branch during these
observations (see \markcite{HVK89}Hasinger \& van der Klis 1989). 
We will present more detailed analysis of the accretion driven flux and
properties of all the X-ray bursts observed in a subsequent paper. 
Figure 1 shows the 2 - 90 keV PCA lightcurve (top) and the hardness ratio 
(6 - 18 keV)/(2 - 6 keV) at 1/16 s resolution for this burst. The drop in the
hardness ratio at about 1.5 s is a clear indication of photospheric radius
expansion. This burst had a peak flux in the 2 - 60 keV band of $7.2 \times 
10^{-8}$ ergs cm$^{-2}$ s$^{-1}$ and a fluence in the same band of $3.9 \times
10^{-7}$ ergs cm$^{-2}$. At the system's probable distance of 6.5 kpc (see
\markcite{L83}Lawrence {\it et al.} 1983) these correspond to $3.6 \times
10^{38}$ ergs s$^{-1}$ and $2.0 \times 10^{39}$ ergs.
We used the 2 - 90 keV, 125 $\mu$s binned data to compute a dynamic
power spectrum by calculating power spectra from 2 s of data every 1/8 s through
the burst. Since the data overlap in this analysis the individual power spectra
are not independent, however, this procedure provides a straightforward way to
investigate the time evolution of the oscillations. The resulting dynamic power
spectrum is presented in figure 2, which shows contours of constant power
spectral amplitude along with the burst profile superposed. This analysis
reveals the presence of strong oscillations at about 580 Hz on the rising edge
of the burst as well as oscillations after photospheric touchdown at a frequency
about 1 Hz higher. As we discuss below, the oscillations during the rise are
transient, lasting about 0.25 s, this introduces a characteristic width to the
peak in the power spectrum which results in the broader distribution in 
frequency of the power spectral contours near the burst rise. We emphasize that
the contours centered near 580.5 Hz at about 2 s are not indicative of a 
real drop in the oscillation frequency at the burst onset. However, the change
in frequency from burst rise to post-peak is real and is not caused by finite
sampling time effects or the transient nature of the oscillations.
This trend for the oscillation frequency to increase with time during some
bursts has now been reported in all sources with millisecond burst oscillations
except KS1731-26, from which only a single burst with oscillations has so far
been detected (see \markcite{Stroh97}Strohmayer {\it et al.} 1997;
\markcite{SMB}Smith, Morgan, \& Bradt 1997).

To further characterize the oscillations seen during the rising phase we
performed an epoch folding period search on the rising interval. Figure 3 shows
the lightcurve of the burst, focusing on the rising portion. The interval used
for this analysis is denoted by the dashed lines in figure 3. The epoch folding
reveals a strong oscillation at 1.723 ms. We folded the data on this period to
determine the
pulse profile and amplitude. The resulting pulse profile is shown in figure 4a.
We fit a model with a constant countrate plus a sinusoid, $A + B\sin(2\pi t/P +
\phi)$, to the pulse profile and find an amplitude $B/A = 30 \pm 4 \%$ with an
acceptable fit.
Since there is no evidence that the persistent flux component is oscillating at
the frequency observed during the burst, to determine the pulsed amplitude we
first subtracted from the folded lightcurve the average persistent countrate
calculated from a 20 s interval prior to the burst. We note that there are no
detectable oscillations during the interval when the drop in hardness ratio at
about 1.5 s (see figure 1) indicates the onset of photospheric radius expansion.
This is consistent with the idea that the expansion of the photosphere 
destroys the coherent oscillations, probably both by scattering away the
amplitude as well as changing the frequency quickly via the expansion of the 
surface layers. After photospheric touchdown (see figure 1 at about 3.5 s) 
the oscillations are again detected, but with slightly higher frequency and 
a much lower amplitude of about 5.7 \% in the 2 - 60 keV bandpass.

The presence of large amplitude oscillations very near the onset of this burst
is similar to the behavior of bursts from 4U 1728-34 reported by
\markcite{SZS}Strohmayer, Zhang \& Swank (1997). We found even higher
oscillation amplitudes by folding a 62 ms interval close to the onset of the
burst. The folded interval is denoted by the solid vertical lines in figure 3.
The background subtracted folded lightcurve obtained from this 62 ms interval 
is shown in figure 4b. We again fit the sinusoidal model to this lightcurve and
found an amplitude of $75 \pm 17 \%$ for this interval. 
This is easily the largest amplitude of the 581 Hz oscillation seen during 
this burst. Such a large amplitude--almost the entire thermal burst flux is
modulated--only about 0.1 s after the onset of the burst is strong evidence 
that we are seeing a rotationally modulated, thermonuclear hotspot.  

\markcite{SZS}Strohmayer, Zhang \& Swank (1997) reported spectral evidence for
an increasing X-ray emitting area during burst rise using bursts from 4U 1728-34
observed with RXTE. We performed similar spectral analysis in order to 
determine whether the same process is plausible for this burst as an 
explanation for the strong oscillations seen during the rise. 
Using the 125 $\mu$s, 64 channel data we 
accumulated spectra through the burst on either 0.0625, 0.125, 0.25 or 0.5 s
intervals depending on the countrate. We also accumulated a 20 s interval just
prior to the burst for use as a background estimate. We fit the background
subtracted spectra to a blackbody model to determine temperatures and fluxes as
a function of time through the burst. We created a plot of $F^{1/4}_{bol} /
kT_{BB}$ versus $F_{bol}$, where $F_{bol}$ is the bolometric flux and $kT_{BB}$
is the blackbody temperature in keV. For a true blackbody the quantity
$F^{1/4}_{bol} / kT_{BB} \propto A^{1/4}/d^{1/2}$, where $A$ is the X-ray
emitting area and $d$ is the distance to the source. We found results very
similar to those reported by Strohmayer, Zhang \& Swank (1997) for bursts from
4U 1728-34, that is, $F^{1/4}_{bol} / kT_{BB}$ increases during the rise. This
is strong evidence for an expanding X-ray emission region and thus supports the
spin modulation hypothesis in this burst from 4U 1636-54. We emphasize that
this result suggests an increase in the X-ray emitting area {\it prior} to the
onset of photospheric radius expansion. The onset of photospheric radius
expansion is reflected in the drop in hardness ratio at about 1.5 s in figure 1.
It also corresponds to an increase in the emitting surface area, but it is
caused by {\it radial} expansion of the photosphere due to radiation forces and
not the {\it lateral} propagation of the burning instability. The blackbody 
spectral analysis indicates that the inferred emitting areas were approximately
the same both immediately before and after the photospheric radius expansion.
This result combined with the large difference in oscillation amplitudes
before and after photospheric expansion suggests that the thermonuclear flash
covered the entire surface of the neutron star.

\section{Discussion}

If the millisecond oscillations seen during bursts are in fact due to spin
modulation, then detailed study and modelling of the oscillation amplitudes,
pulse profiles and spectral variability with pulse phase during X-ray bursts can
provide a wealth of information on the mass and radius of the neutron star. For
example, the maximum modulation amplitude that can be obtained from a hotspot of
a given angular size on a rotating neutron star is set by the strength of
general relativistic light bending. For the case that rotation is not rapid
enough to substantially distort
the exterior spacetime from the Schwarzchild spacetime, and this is the case 
even for spin periods of a few milliseconds (Lamb \& Miller 1995),
then the maximum amplitude depends only on the compactness of the neutron star,
that is, the ratio of stellar mass to radius $GM/c^2R$. 
Stars which are more compact produce lower
amplitudes due to flattening of the pulse by light bending 
(see \markcite{PFC}Pechenick, Ftaclas \& Cohen 1983; \markcite{S92}Strohmayer
1992; and \markcite{ML97}Miller \& Lamb 1997). Since the intrinsic rotational
modulation amplitude can only be decreased by other effects such as photon
scattering (see \markcite{MLP}Miller, Lamb \& Psaltis 1997;
\markcite{BL}Brainerd \& Lamb 1987; and \markcite{KP}Kylafis \& Phinney 1989) 
or the viewing geometry of the spot, the maximum observed oscillation amplitude
represents a lower limit to the intrinsic amplitude. 
Thus an observed amplitude can be used to place an upper limit on the
compactness of the neutron star, that is, if the star were more compact than
some limit it would not be able to produce a modulation amplitude as large as
that observed. 

In principle, stellar rotation will also play a role in the observed properties
of spin modulation pulsations. For example, assuming the oscillation frequency
of 581 Hz represents the spin frequency of the neutron star in 4U 1636-54, then
for a 10 km radius neutron star the spin velocity is $v_{spin}/c = 2\pi
\nu_{spin} R  \approx 0.12$ at the rotational equator. The motion of
the hotspot produces a Doppler shift of magnitude $\Delta E / E \approx
v_{spin}/c = 0.12$, thus the observed spectrum is a function of pulse phase 
(see \markcite{CS}Chen \& Shaham 1989). Measurement of a pulse phase dependent
Doppler shift in the
X-ray spectrum would provide additional evidence supporting the spin modulation
model and would also provide a means of constraining the neutron star radius.
The rotationally induced velocity also produces an aberration which 
results in asymmetric pulses, thus the pulse shapes also contain information on
the spin velocity and therefore the stellar radius 
(\markcite{CS}Chen \& Shaham 1989). The component of the spin velocity 
along the line of site is proportional to $\cos\theta$, where $\theta$ is the
lattitude of the hotspot measured with respect to the rotational equator. The
modulation amplitude also depends on the lattitude of the hotspot, as spots near
the rotational poles produce smaller amplitudes than those at the equator. Thus
we expect a correlation between the observed oscillation amplitude and the
size of any pulse phase dependent Doppler shift. Dectection of such a
correlation in a sample of bursts would definitively confirm the rotational
modulation model in our opinion. We will present calculations of mass - radius
constraints for 4U 1636-54 and 4U 1728-34 including the effects of light
bending, rotation and angle dependent emission, based on the observed properties
of burst oscillations as well as spectroscopy of Eddington limited bursts in a
subsequent paper.

Several models for the kilohertz QPO seen in 13 LMXB systems (see
\markcite{vdk}van der Klis 1997 for a recent review) invoke some sort of
beat-frequency interpretation for the twin kHz peaks seen in many of the sources
(see \markcite{MLP}Miller, Lamb \& Psaltis 1997;
\markcite{S96}Strohmayer {\it et al.} 1996). So far only in 4U 1728-34 does the
frequency difference between the twin kHz peaks match the frequency observed
during X-ray bursts (see \markcite{S96}Strohmayer {\it et al.} 1996). In two
other sources
(KS 1731-26 and 4U 1636-54) the separation of the twin kHz peaks is closer to
1/2 the frequency of oscillations observed during bursts (see
\markcite{WV}Wijnands \& van der Klis 1997; and \markcite{Z96}Zhang {\it et al.}
1996). For example, \markcite{W97}Wijnands {\it et al.} (1997) report a
frequency difference for the twin kHz QPO in 4U1636-54 of $276 \pm 10$ Hz. 
The effort to reconcile
these observations with a beat-frequency interpretation has led to speculation
that the oscillation frequency observed during bursts may sometimes be twice the
spin frequency of the star, although in 4U1636-54 the difference frequency 
appears to be a bit less than 1/2 the burst oscillation frequency, and in some
Z sources the frequency difference is not constant (\markcite{van97}van der Klis
{\it et al.} 1997). If this scenario is
correct it implies the existence of two antipodal spots on the neutron star
surface during X-ray bursts. In addition to the daunting requirement of 
initiating the thermonuclear flash nearly simultaneously on opposite sides of
the neutron star, the observation of large oscillation amplitudes
shortly after burst onset in 4U1636-54 places severe constraints on the two
hotspot scenario. To see this one can ask the following question. What maximum
amplitude can be produced by a star with antipodal hotspots? For two antipodal
spots light bending strongly constrains the amplitudes that can be achieved
(see \markcite{PFC}Pechenick, Ftaclas \& Cohen 1983). Calculations using the
Schwarzchild spacetime and isotropic emission from the stellar surface indicate
that even a neutron star with an implausibly small compactness of $M/R = 0.1$,
recall that rotational modulation amplitude increases with decreasing
compactness, can only achieve a maximum amplitude of about 30 \%, whereas we
measured an amplitude of 75\% from the burst described here. We note that with
$M/R = 0.1$ a 1.4 $M_{sun}$ neutron star would
have a radius of 21 km. This is far stiffer than any neutron star equation of 
state that we are familiar with. Rapid rotation could in principle modify this
result, but at the modest inferred spin period of 290.5 Hz for the two spots to
produce the observed 581 Hz frequency one would still require implausibly large
neutron star radii to make a significant rotational correction to the amplitude.
Another process which can increase the amplitude is beaming of radiation at the
stellar surface. For example, to the extent that electron scattering is the
dominant opacity process in neutron star atmospheres then one should expect a
specific intensity distribution which approximates that from a grey atmosphere.
Such a distribution is proportional to $\cos\delta + 2/3$, where $\delta$ is the
angle from the normal to the stellar surface (see \markcite{Mihal}Mihalas 1978),
so this will modestly increase the rotational modulation amplitude. In addition,
it is likely that the emergent spectrum will also be a weak function of 
$\delta$ (\markcite{ML97}Miller \& Lamb 1997). With more detailed modelling of
the emission from hotspots of finite size it will be possible to place strong
constraints on the antipodal hotspot interpretation for the burst oscillations.
Finally, if further analysis such as pulse phase spectroscopy continues to
support the interpretation of the burst oscillation frequency as the neutron
star spin frequency in 4U1636-54, then kHz QPO frequency separations near 
1/2 the spin frequency would suggest that the beat frequency interpretation 
may be untenable.
 
\acknowledgements

We thank Lars Bildsten, Andrew Cumming, Keith Jahoda, and John Wang for
helpful discussions. We thank the anonymous referee for helping us improve the
manuscript. T. S. acknowledges the High Energy Astrophysics Program
under USRA in the Laboratory for High Energy Astrophysics (LHEA) at NASA's
Goddard Space Flight Center.

\vfill\eject

\vfill\eject

\section{Figure Captions}

\noindent Figure 1: 2 - 90 keV lightcurve and hardness ratio plot for the burst.
The hardness ratio is defined as the ratio of counts in the 6 - 18 keV range to
that in the 2 - 6 keV range. The drop in hardness ratio at about 1.5 s indicates
the beginning of the radius expansion phase. The dashed vertical lines denote
the time interval used in the epoch folding analysis.

\vskip 10pt

\noindent Figure 2: Representation of the dynamic power spectrum for the burst
shown in figure 1. The contours are levels of constant power spectral amplitude.
The power spectra were calculated from 2 s intervals with the center of each 
succesive interval shifted by 1/8 s. The countrate profile of burst 1 is also
shown. The burst profile is aligned to the center of each data interval used
to calculate the power spectra.
 
\vskip 10pt

\noindent Figure 3: Lightcurve of the burst focusing on the rise. The 
interval denoted by dashed lines is that used for the folding analysis shown in
figure 4a, while the solid vertical lines denote the 62.5 ms interval used for
the folding analysis displayed in figure 4b.

\vskip 10pt

\noindent Figure 4: (a) Folded, background subtracted lightcurve using the 0.25
second interval denoted with dashed vertcal lines in figure 3. The best fitting
sinusoidal model is also shown. The average amplitude during this interval was
$30 \pm 4 \%$. (b) The same results are shown for the 62.5 ms interval shown in
figure 3. For this interval the amplitude was $75 \pm 17\%$.

\vskip 10pt

\vfill\eject

\begin{figure*}[htb] 
\centerline{\epsfig{file=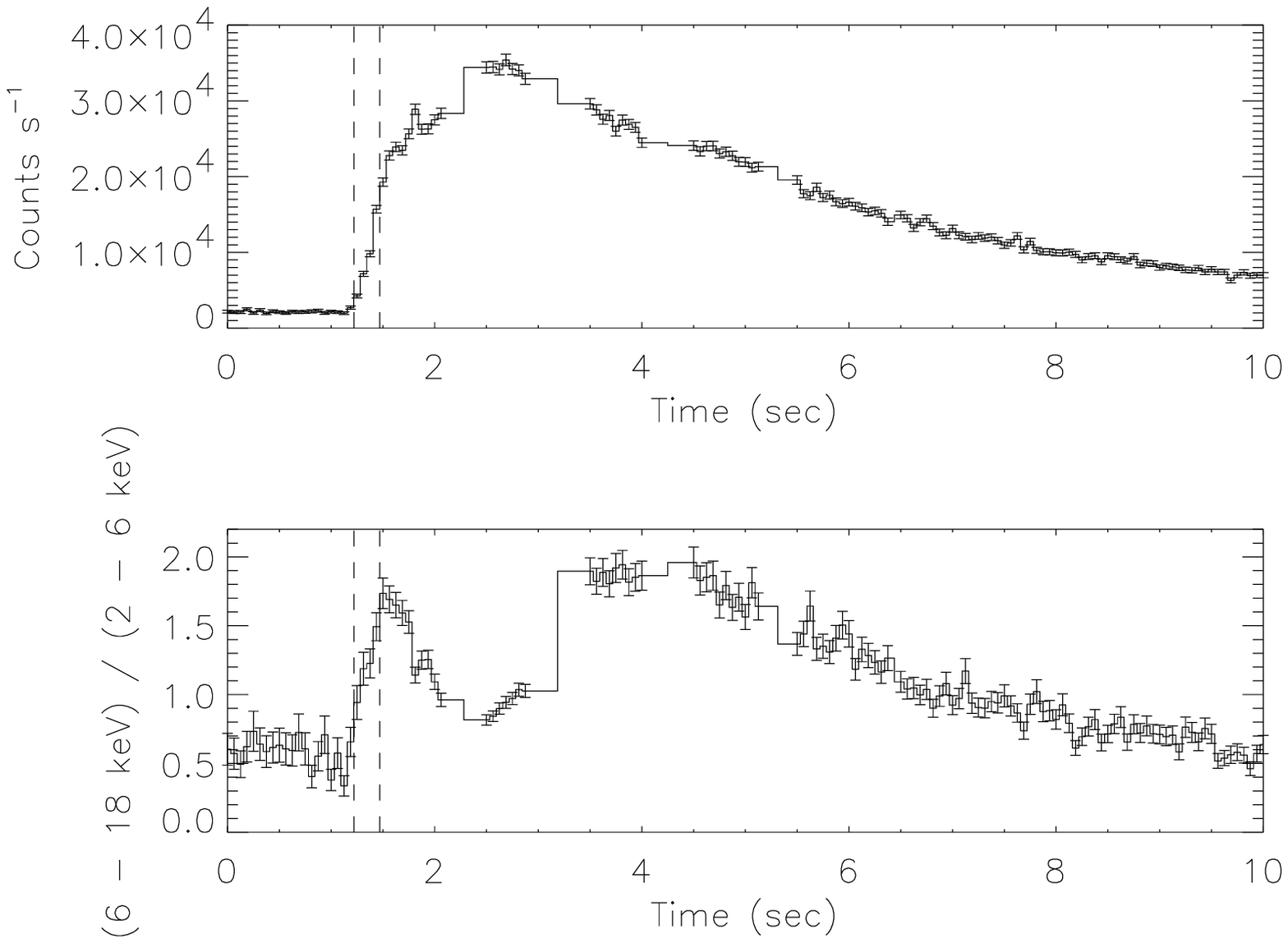,height=5.5in,width=5.5in}}
\vspace{10pt}
\caption{Figure 1}
\label{fig1}
\end{figure*}

\vfill\eject

\begin{figure*}[htb] 
\centerline{\epsfig{file=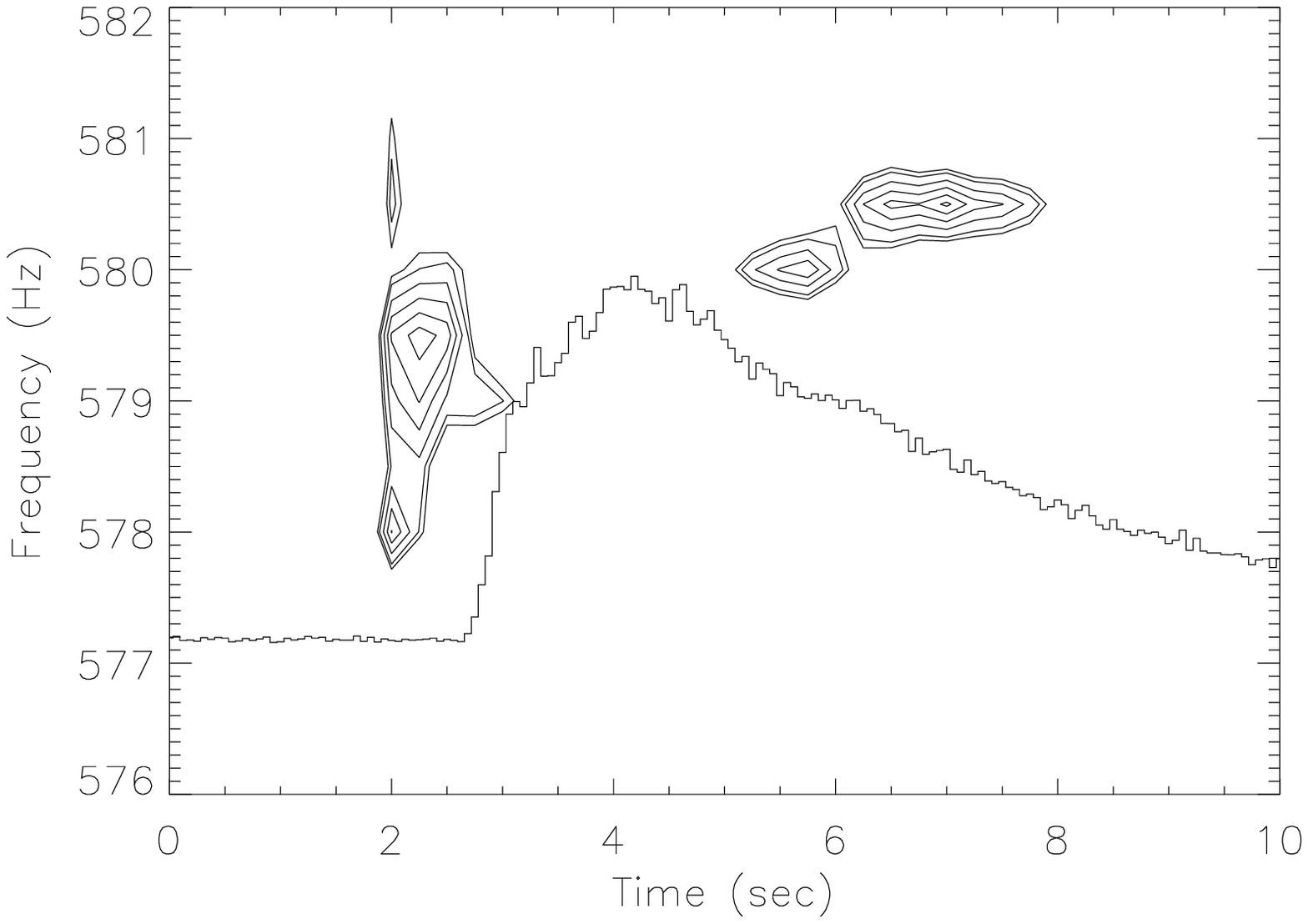,height=5.5in,width=5.5in}}
\vspace{10pt}
\caption{Figure 2}
\label{fig2}
\end{figure*}

\vfill\eject

\begin{figure*}[htb] 
\centerline{\epsfig{file=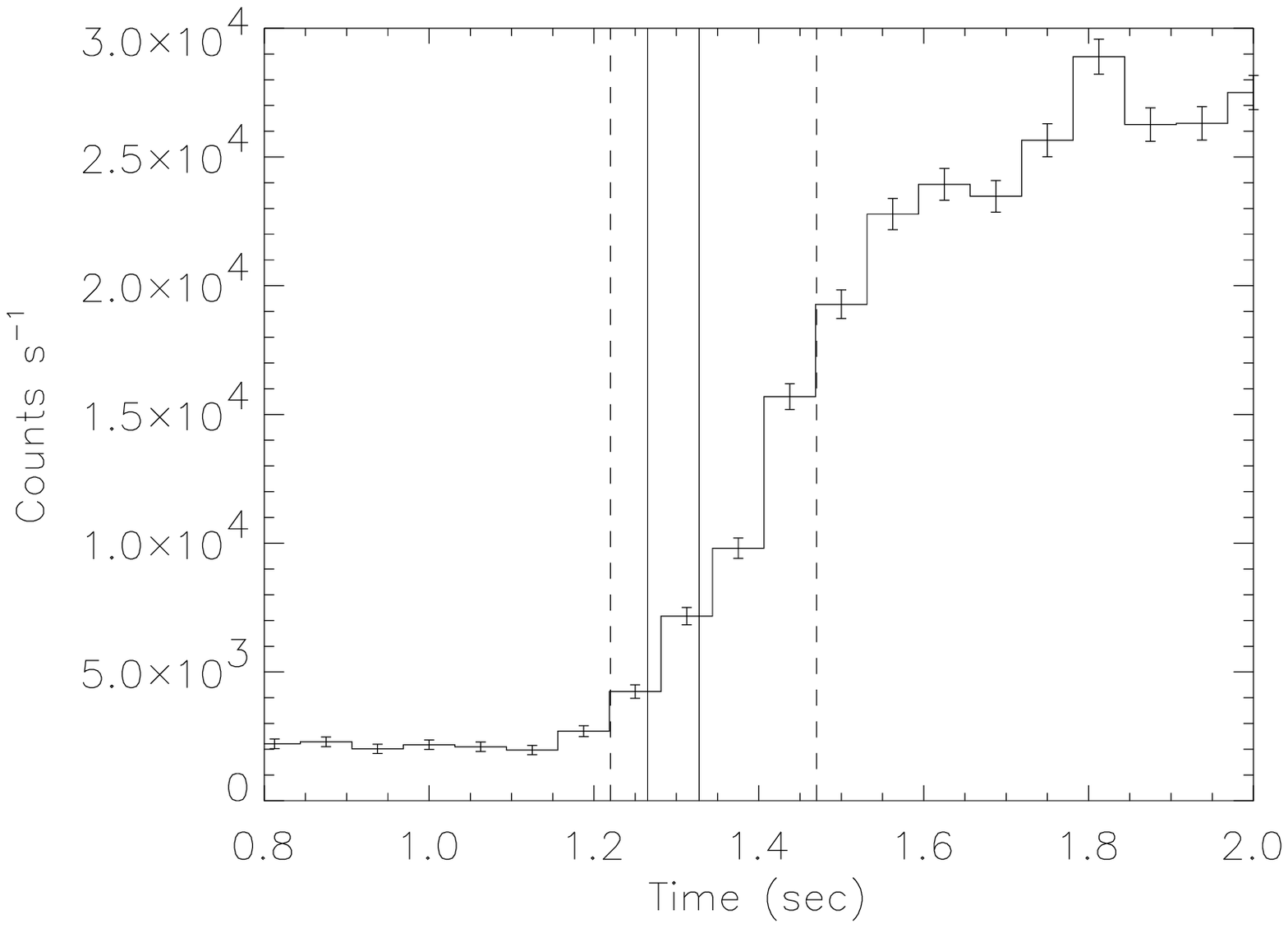,height=5.5in,width=5.5in}}
\vspace{10pt}
\caption{Figure 3}
\label{fig3}
\end{figure*}

\vfill\eject

\begin{figure*}[htb] 
\centerline{\epsfig{file=brise_1636_250ms_flc_fig.ps,height=5.5in,width=5.5in}}
\vspace{10pt}
\caption{Figure 4a}
\label{fig4a}
\end{figure*}

\vfill\eject

\begin{figure*}[htb] 
\centerline{\epsfig{file=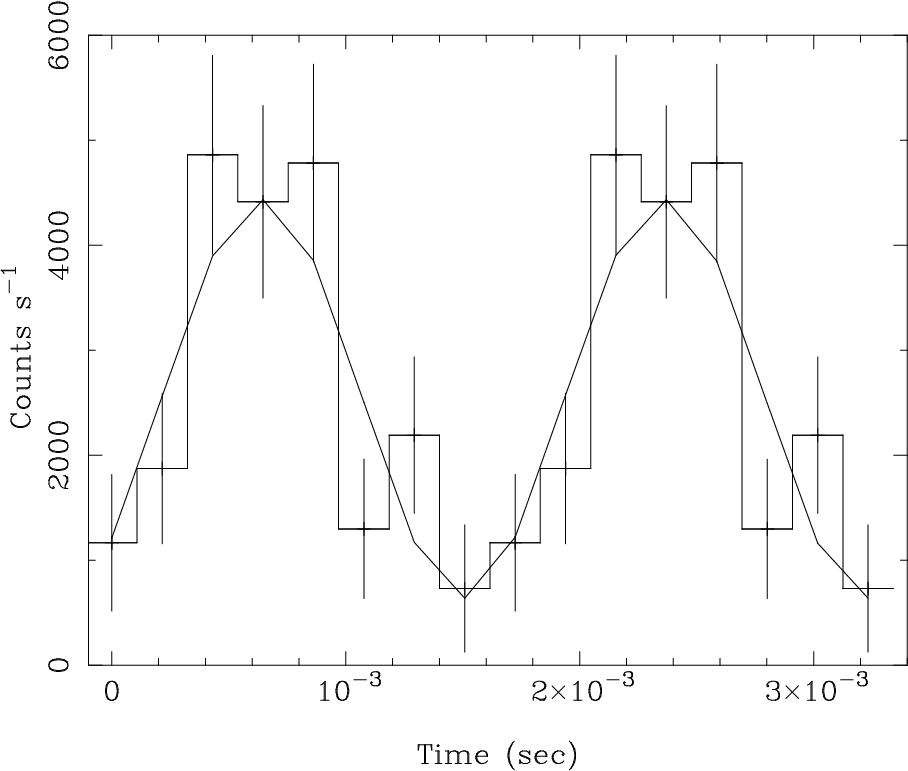,height=5.5in,width=5.5in}}
\vspace{10pt}
\caption{Figure 4b}
\end{figure*}


\begin{references}

\reference{B98} Bildsten, L. 1997, to appear in Proc. NATO ASI "The Many
Faces of Neutron Stars", Lipari, Italy, 1996

\reference{B95} Bildsten, L. 1995, ApJ, 438, 852

\reference{BL87} Brainerd, J. J., \& Lamb, F. K. 1987, ApJ, 317, L33

\reference{CS} Chen, K. \& Shaham, J. 1989, ApJ, 339, 279

\reference{FW82} Fryxell, B. A., \& Woosley, S. E. 1982, ApJ, 261, 332

\reference{HVK89} Hasinger, G. \& van der Klis, M. 1989, A\& A, 225, 79

\reference{Joss78} Joss, P. C. 1978, ApJ, 225, L123

\reference{KP89} Kylafis, N. \& Phinney, E. S. 1989, in Timing Neutron Stars, 
eds. H. Ogelman and E. P. J. van den Heuvel, (Dordrecht:Kluwer), p. 731

\reference{L83} Lawrence, A. {\it et al.} 1983, ApJ, 271, 793

\reference{Mihal} Mihalis, D. 1978, in Stellar Atmospheres, (W. H. Freeman)

\reference{ML97} Miller, M. C. \& Lamb, F. K. 1997, ApJ submitted.

\reference{MLP} Miller, M. C., Lamb, F. K., \& Psaltis, D. 1997, ApJ, to appear

\reference{NIF84} Nozakura, T., Ikeuchi, S. \& Fujimoto, M. Y. 1984, ApJ, 286,
221

\reference{PFC} Pechenick, K. R., Ftaclas, C., \& Cohen, J. M. 1983, ApJ, 275,
846

\reference{SMB} Smith, D., Morgan, E. H. \& Bradt, H. V. 1997, ApJ, in press.

\reference{SZS} Strohmayer, T. E., Zhang, W. \& Swank, J. H. 1997, ApJ, 487, L77

\reference{S2} Strohmayer, T. E. 1992, ApJ, 388, 138

\reference{S96} Strohmayer, T. E., Zhang, W., Swank, J. H., Smale, A. P.,
Titarchuk, L., Day, C. \& Lee, U. 1996, ApJ, 469, L9

\reference{Stroh97} Strohmayer, T. E., Jahoda, K., Giles, A. B. \& Lee, U. 1997,
ApJ, 486, 355

\reference{Swank97} Swank, J. H. {\it et al.} 1997, in preparation

\reference{van97} van der Klis, M., Wijnands, R., Chen, W. \& Horne, K. 1997 
ApJ, 481, L97 {\it et al.} 1997, 

\reference{VDK} van der Klis, M. 1997, to appear in Proc. NATO ASI "The Many
Faces of Neutron Stars", Lipari, Italy, 1996.

\reference{W97} Wijnands, R. {\it et al.} 1997, 479, L141

\reference{WV} Wijnands, R. A. D., \& van der Klis, M. 1997, ApJ, 482, L65

\reference{Z96} Zhang, W., Lapidus, I., Swank, J. H., White, N. E. \&
Titarchuk, L. 1996, IAUC 6541

\end{references}
\end{document}